\documentclass[fleqn,usenatbib]{mnras}

\usepackage{newtxtext,newtxmath}
\usepackage[T1]{fontenc}

\DeclareRobustCommand{\VAN}[3]{#2}
\let\VANthebibliography\thebibliography
\def\thebibliography{\DeclareRobustCommand{\VAN}[3]{##3}\VANthebibliography}

\usepackage{graphicx}	% Including figure files
\usepackage{amsmath}	% Advanced maths commands
\usepackage{xcolor}
\usepackage{subcaption}
\newcommand{\red}{\textcolor{black}}

\title[Confirming Planetary Trends]{Confirming known planetary trends using a photometrically selected \textit{Kepler} sample}

\author[Hansen et al.]{
Jonah T. Hansen,$^{1}$\thanks{E-mail: jonah.hansen@anu.edu.au}
Luca Casagrande,$^{1}$
Michael J. Ireland$^{1}$
and Jane Lin$^{1}$
\\
$^{1}$Research School of Astronomy and Astrophysics, The Australian National University, Canberra, ACT 2611, Australia\\
}

\date{Accepted XXX. Received YYY; in original form ZZZ}

\pubyear{2020}

\begin{document}
\label{firstpage}
\pagerange{\pageref{firstpage}--\pageref{lastpage}}
\maketitle

% Abstract of the paper
\begin{abstract}
Statistical studies of exoplanets and the properties of their host stars have been critical to informing models of planet formation. Numerous trends have arisen in particular from the rich \textit{Kepler} dataset, including that exoplanets are more likely to be found around stars with a high metallicity and the presence of a ``gap'' in the distribution of planetary radii at 1.9\,$R_\oplus$. Here we present a new analysis on the \textit{Kepler} field, using the APOGEE spectroscopic survey to build a metallicity calibration based on \textit{Gaia}, 2MASS and Str\"omgren photometry. This calibration, along with masses and radii derived from a Bayesian isochrone fitting algorithm, is used to test a number of these trends with unbiased, photometrically derived parameters, albeit with a smaller sample size in comparison to recent studies. We \red{recover} that planets are more frequently found around higher metallicity stars; over the entire sample, planetary frequencies are $0.88\pm0.12$~percent for [Fe/H]\,<\,0 and $1.37\pm0.16$~percent for [Fe/H]\,$\geq$\,0 \red{but at two sigma we find that the size of exoplanets influences the strength of this trend}. We also recover the planet radius gap, along with a slight positive correlation with stellar mass. We conclude that this method shows promise to derive robust statistics of exoplanets. We also remark that spectrophotometry from \textit{Gaia} DR3 will have an effective resolution similar to narrow band filters and allow to overcome the small sample size inherent in this study. 
\end{abstract}

% Select between one and six entries from the list of approved keywords.
% Don't make up new ones.

%%%%%%%%%%%%%%%%%%%%%%%%%%%%%%%%%%%%%%%%%%%%%%%%%%

%%%%%%%%%%%%%%%%% BODY OF PAPER %%%%%%%%%%%%%%%%%%

\section{Introduction}
\label{sec:intro}
Since the release of \textit{Kepler}'s \citep{Kepler} rich collection of over 4000 exoplanet transits, the study of exoplanet statistics has blossomed into a thriving field. Exoplanet demographic studies have led to many interesting results, including the preference of large exoplanets ($R_p > 4\,R_\oplus$) to form around high metallicity stars \citep{santos_spectroscopic_2004,fischer_planet-metallicity_2005,zhu_influence_2019} and that close, multiple planet systems form preferentially around metal poor stars \citep{brewer_compact_2018}. However, perhaps one of the most important results to have come out of these studies it that of the planet radius ``gap'' - a decrease in the number of planets with radii around 1.5-2.0\,$R_\oplus$ \citep[e.g.,][]{owen_kepler_2013,fulton_california-kepler_2017}. This particular radius is significant, as it separates the classification of ``super-Earths'' from ``sub-Neptunes''. Reasons for the presence of this gap are numerous, ranging from UV photoevaporation of a planet's atmosphere \citep{owen_kepler_2013,owen_evaporation_2017,lopez_how_2018} to core-powered mass loss \citep{ginzburg_core-powered_2018,gupta_signatures_2020}. 

Recently, more separate studies have found evidence for this gap, using both the \textit{Kepler} \citep{van_eylen_asteroseismic_2018,berger_gaia-kepler_2020-2} and K2 surveys \citep{hardegree-ullman_scaling_2020,cloutier_evolution_2020} and a handful have also identified the gap follows a trend with stellar mass: the drop in occurrence is found at smaller radii around less massive stars \citep{fulton_california-kepler_2018,berger_gaia-kepler_2020-2}. The radius gap still has ambiguity in the strength of this deficiency, with studies such as the seminal \citet{fulton_california-kepler_2017} revealing a shallow gap, whereas \citet{van_eylen_asteroseismic_2018}, with a much smaller sample but with very precise parameters from asteroseismology, find a gap nearly devoid of planets. 
\red{These differences are thoroughly discussed in \citet{petigura_two_2020}, who shows how the depth of the radius gap is sensitive to various sample cuts, highlighting the challenges of performing population statistics.}

Part of the ambiguity around these planetary trends, including the radius gap, stem from the imprecise nature of the \textit{Kepler} input catalogue (KIC). The KIC was a compilation of 13 million stars with optical photometry and stellar parameters created for the purpose of choosing \textit{Kepler} targets, of which approximately 200,000 were chosen \citep{batalha_selection_2010,KIC}. However, the parameters for these stars were lacking in precision and some critical parameters, such as the age and mass of these stars, were missing entirely. To investigate exoplanet demographics, precise stellar parameters are required which has led to many follow-up studies of the stars in the KIC \citep[e.g.,][]{bruntt,molenda,huber_revised_2014,petigura,furlan_kepler_2018,berger_gaia-kepler_2020-1}. Many of these studies rely on spectroscopy, for which \red{selection effects can be stronger and more difficult to quantify than in a photometrically selected sample of stars. This point will be discussed in more detail in Section \ref{sec:representative}.}

In this paper, we derive stellar parameters for a photometrically unbiased sample of confirmed \textit{Kepler} transiting planets to study some of the known trends concerning exoplanetary demographics. We assemble our sample starting from the Str\"omgren survey for Asteroseismology and Galactic Archaeology \citep[SAGA,][]{stromgren} and complementing it with photometry from \textit{Gaia} DR2 \citep{Gaia} and 2MASS \citep{skrutskie_two_2006}.
 Metallicity of the \textit{Kepler} host stars is derived through a photometric calibration based on the APOGEE survey \citep{apogee}, and effective temperatures are calculated through the Infra-Red Flux Method \citep[IFRM; e.g.,][]{black77,c10}  We then calculate the masses and radii of \textit{Kepler} host stars through \textit{Gaia} parallaxes and isochrone fitting by using the Bayesian isochrone fitting algorithm \textsc{Elli} \citep{lin_stellar_2018}. This results in us obtaining a similar planet radius-mass trend to that of \citet{fulton_california-kepler_2018} and \citet{berger_gaia-kepler_2020-2}, as well as finding large planets preferentially form around metal rich stars and a slight trend that multiple exoplanet systems form around metal poor stars. 

\section{Catalogue Compilation}
\label{sec:catalogue}
Multiple stellar catalogues were combined to leverage finding an appropriate metallicity index for our planet host star sample. Foremost of these was the aforementioned \textit{Kepler} Input Catalogue (KIC), a catalogue of stars that lie in the \textit{Kepler} field \citep{KIC}. Not all the stars present in the KIC contain useful data on their properties and so a subset of the catalogue was used: all KIC objects viewed in Quarter 15 of the \textit{Kepler} mission that had long cadence data. 

The KIC catalogue was matched with the \textit{Gaia} DR2 catalogue \citep{Gaia} to obtain \textit{Gaia} photometry and parallaxes for these stars. We remark that for the isochrone fitting described in Section \ref{sec:radii} we use distances from \citet{Bailer-Jones2018}. The \textit{Gaia} data for these KIC stars was combined with other photometric catalogues, including the 2MASS catalogue's $J$, $H$ and $K$ band photometry and the Str\"omgren catalogue's $u$, $v$, $b$ and $y$ band photometry produced by \cite{stromgren}. These catalogues were cross-matched so that all stars contained the photometry from each survey; resulting in our catalogue of multi-band photometry encompassing around 30,000 stars. We note here that this is a small fraction of the KIC, primarily due to to the small fraction of stars in the \textit{Kepler} field currently with Str\"omgren photometry.

The photometry was corrected for reddening using the \cite{reddening} reddening map. This map is known to overestimate reddening along the galactic plane \citep[see e.g.,][]{Arce,Schlafly}. Hence, it was re-scaled by the following formula where $b$ is the galactic latitude:
\begin{equation}
    E(B-V)_\text{res} = E(B-V) +0.1\log(|b|-3)-0.16
\end{equation}
which is appropriate for the range $5 \lesssim b \lesssim 20$ encompassed by the \textit{Kepler} field, and whose derivation 
is explained in \cite{reddening_rescale} and \cite{skymapper}. Magnitudes were de-reddened using extinction coefficients from \citet{extinction1,extinction2}. 

The photometric stellar catalogue was finally combined with the list of \textit{Kepler} Objects of Interest (KOI). This is a list of all candidate exoplanets found in the \textit{Kepler} field, provided by the NASA Exoplanet Archive (\url{https://exoplanetarchive.ipac.caltech.edu/}). Objects with a \textsc{koi disposition} flag of \textsc{false positive} were removed and the remaining entries were paired with the photometric data from their host star, producing a separate KOI catalogue of about 800 exoplanets and their host stars. 

\section{Metallicity Calibration}
\label{sec:metallicity}
In order to derive homogeneous metallicities for all stars in our catalogue, we devise a metallicity calibration using the photometry assembled in Section \ref{sec:catalogue}. The largest sample of stars with spectroscopic metallicities in the \textit{Kepler} field is from APOGEE \citep[APO Galactic Evolution Experiment; e.g.,][]{apogee}, an infrared spectroscopic mission that was designed to measure the radial velocities and, more importantly, chemical abundances of over 100,000 red giants \red{within} the Milky Way. We used the [M/H] metallicity from APOGEE DR14 \citep{apogee_DR14} to calibrate our photometry; removing stars with the \textsc{star\_bad} label and combining this data with the photometric catalogue compiled above to obtain a total of 2415 stars in the KIC with known metallicities. These stars were then used to derive a metallicity relation for the KIC as a whole. 

To derive the best relation a \red{Principal} Component Analysis (hereafter referred to as PCA) decomposition was performed over 84 linear combinations of colour indices, \red{allowing us to reduce the dimensionality of the data to arrive at the best proxy for metallicity.} \red{This number of colour indicies was arrived at by taking six indices that were suggested to be sensitive to metallicity, including the well established  $m_1 = (v-b)-(b-y)$ from the Str\"omgren photometric system, as well as $u-b$, $G-K$, $v-G$, $R_p - K$, and $(B_p-R_p) - (R_p - K)$. Then, a set of all 3-combinations of these indices with repetitions, including the absence of an index, was gathered. That is, all multisubsets of size 3 (allowing up to cubic powers) from the set $\{1,m_1,u-b,G-K,v-G,R_p-K,(B_p-R_p)-(R_p-K)\}$. This provides us with the 84 index combinations described.}

\begin{figure}
  \centering
  \includegraphics[width=\linewidth]{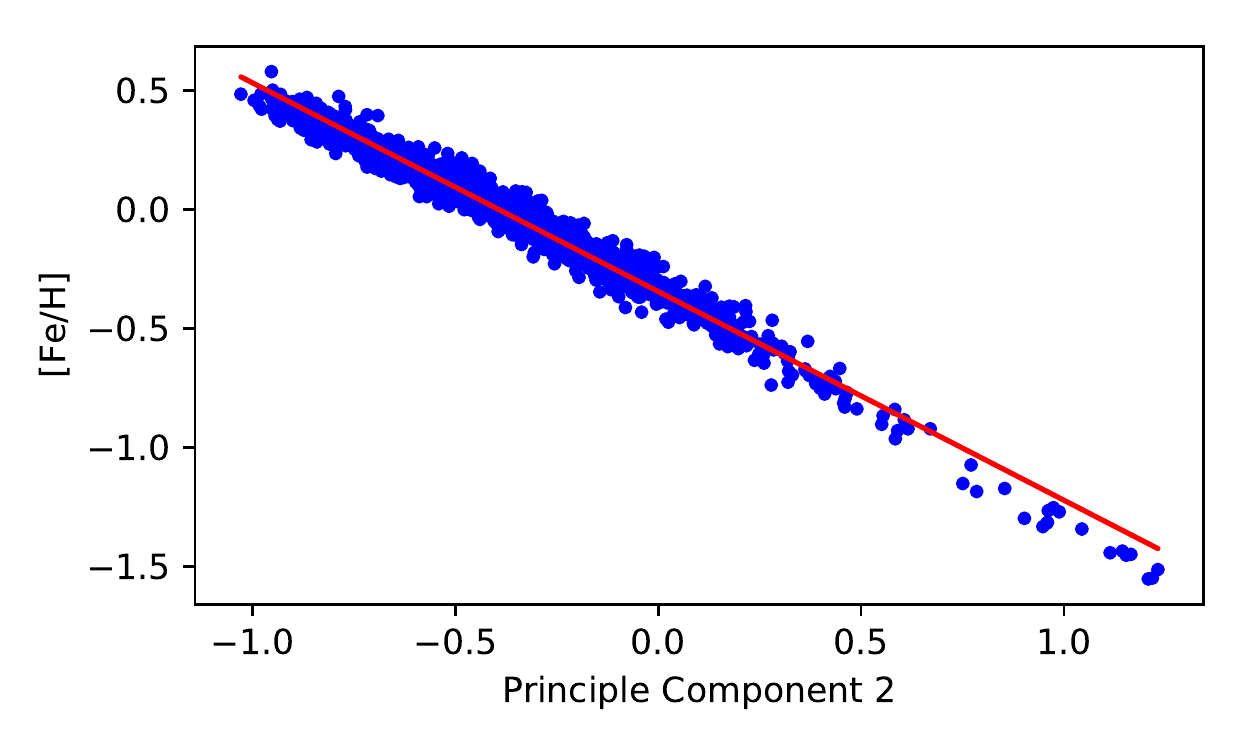}
  \caption{Metallicity against the second \red{principal} component from the PCA. A linear regression line is shown in red, \red{highlighting that the correlation between the principal component and the APOGEE metallicity is well described by a linear fit.}}
  \label{img_PCA_metal}
\end{figure}

A singular value decomposition was performed on the collection of 84 colour indices for our APOGEE cross-matched photometric catalogue;  outputting linear combinations of the input dimensions called ``\red{Principal} Components''. The first of these vectors describes the linear combination of variables that produces the most variance in the data, the second giving the combination that produces the second most variance and so forth. \red{Broad- and intermediate-band stellar colours like those used in this work depend primarily on the effective temperature of the star, and to a lesser extent onto other quantities such as metallicity and surface gravity \citep[e.g.,][]{bessell2005}. We determined from a tight correlation with the APOGEE metallicity that the second principle acted as a good proxy for this quantity.} The correspondence between the second \red{principal} component and the APOGEE metallicity can be seen in Fig. \ref{img_PCA_metal}. \red{This particular component was of the form of Equation 2, which will be addressed in more detail in the following discussion, and shows a linear relationship to the APOGEE metallicity. We note a very slight deviation at metallicities below $-1$, but this is not concerning due to most planet hosting stars having larger metallicities than this.}

With the metallicity \red{principal} component identified, we ran an iterative process to reduce the number of input parameters so that the resulting calibration was not over determined. The process is as follows: the decomposition was completed with \red{the 84 colour index combinations addressed earlier, and the combination} with the weakest contribution to the second \red{principal} component was removed. The decomposition was then performed again with $n-1$ dimensions, repeating the process \red{until the tightest correlation with the fewest parameters remained. At this point, removing another colour index contribution would greatly impact the correlation with APOGEE metallicity.}

At the end of this procedure, we found that the best colour indices to use were a linear combination of the $u-b$ index from the Str\"omgren photometric system and the \red{$(G-K)^2$} index combining the $G$ band from \textit{Gaia}'s photometry and the $K$ band from 2MASS.  A second round of PCA was conducted with these two indices, as well as the APOGEE metallicity itself, resulting in a linear calibration of the form
\begin{equation}
    \text{[Fe/H]}_\text{cal} = a_0(G-K)^2 + a_1(u-b) + a_2
\end{equation}

This calibration still had some residual trends, particularly in the $G-K$ colour index. To correct this, we further fitted a 5th order polynomial \red{in $G-K$} to the residuals and subtracted this from the \red{metallicity calibration in Equation 2. This polynomial was derived from adding terms to the residual fit until the trend was flattened by eye within the scatter.} Hence, our final calibration was of the form:
\begin{equation}
\begin{split}
\text{[Fe/H]}_\text{cal} &= c_1(G-K)^5 + c_2(G-K)^4 + c_3(G-K)^3 \\&+ c_4(G-K)^2 + c_5(G-K) + c_6(u-b) + c_7
\end{split}
\end{equation}
with calibration parameters:
\begin{align*}
    c_1 &= -0.346793
    & c_2 &= 3.108458\\
    c_3 &= -10.38798
    & c_4 &= 15.35047\\
    c_5 &= -10.86391
    & c_6 &= 1.729865\\
    c_7 &= 0.875089
\end{align*}

\begin{figure*}
  \centering
  \begin{subfigure}{0.8\textwidth}
    \centering
    \includegraphics[width=\linewidth]{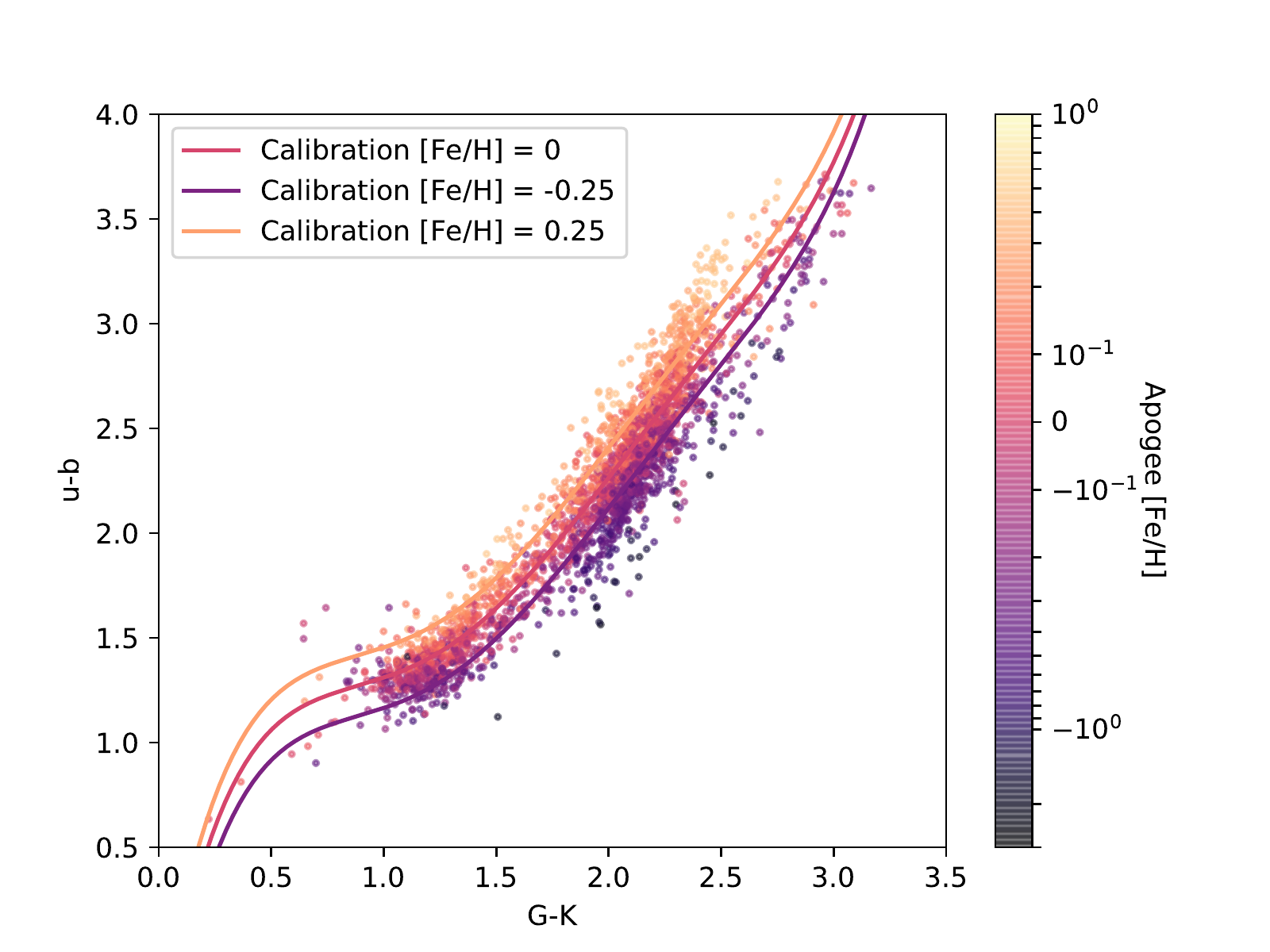}
    \caption{}
    \label{img_callibration_apogee}
  \end{subfigure}
  \begin{subfigure}{0.49\textwidth}%
    \centering
    \includegraphics[width=\linewidth]{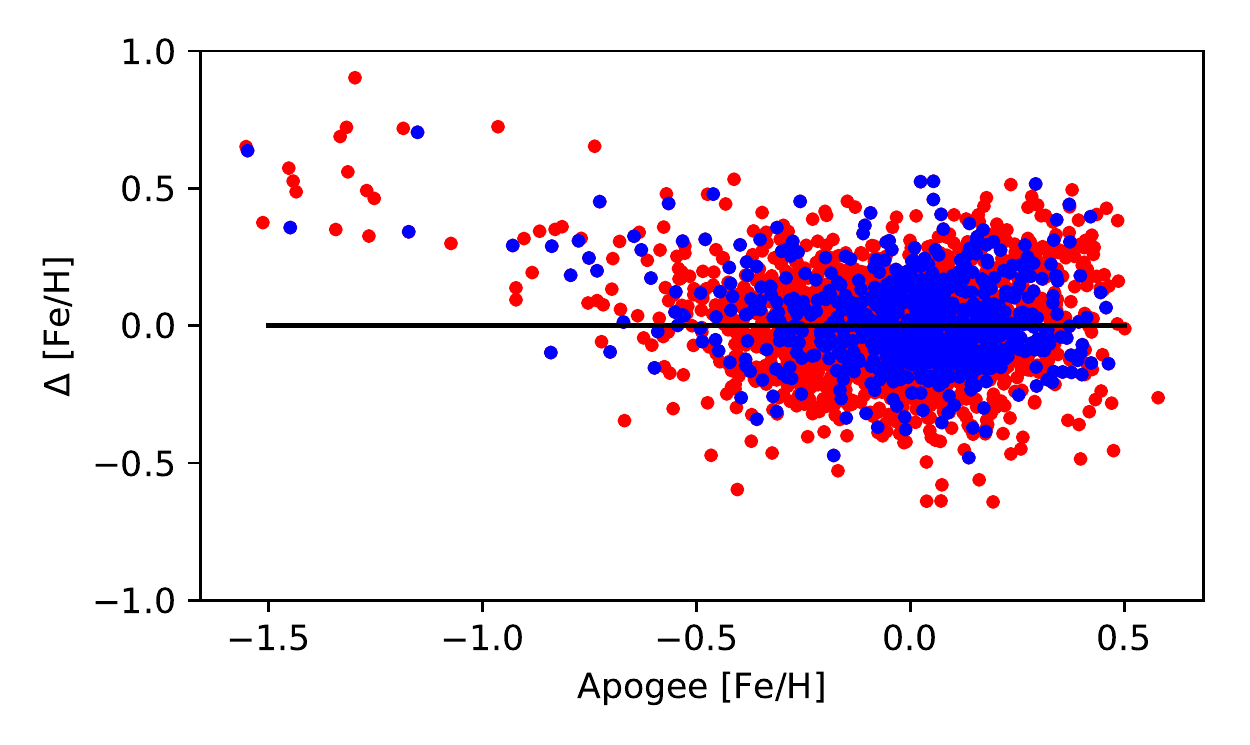}
    \caption{}
    \label{img_res_apogee}
  \end{subfigure}
  \begin{subfigure}{0.49\textwidth}%
    \centering
    \includegraphics[width=\linewidth]{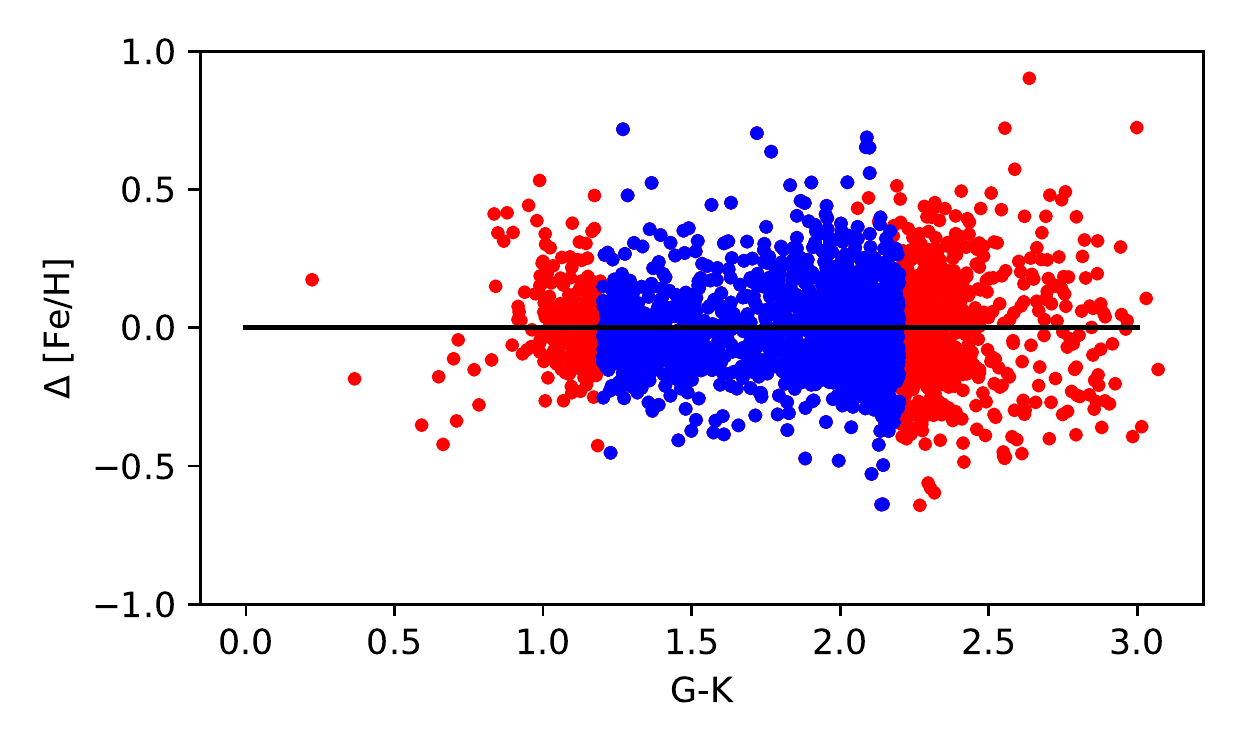}
    \caption{}
    \label{img_res_GK}
  \end{subfigure}
  \caption{Comparison of our photometric metallicity calibration against that of APOGEE. a) The $G-K$, $u-b$ colour plane coloured by the APOGEE [M/H] metallicity index. Our photometric metallicity calibration is shown by the lines for a given metallicity. b) Metallicity residuals (our metallicity - APOGEE) against APOGEE metallicity. Red plots the full catalogue of stars, whereas blue only plots stars that fall within the colour cut discussed in the text. c) Same as b, but against the $G-K$ colour index.}
  \label{img_callibration}
\end{figure*}

Our calibration into the $G-K$ vs $u-b$ plane is shown in the upper panel of Fig. \ref{img_callibration}, where stars are colour coded by their APOGEE metallicity. Also shown in the bottom panels is the residual of our photometric metallicity calibration as function of spectroscopic [Fe/H] and $G-K$. The standard deviation of the residuals (shown in red) is 0.18\,dex. \red{Again we note} that for [Fe/H]~$\lesssim -1$, our calibration systematically overestimates the true metallicity, \red{as seen in the APOGEE [Fe/H] residual plot}. However, this is of little concern for the sake of our study, since the bulk of planets lie well above this limit.

Especially when looking at the $G-K$ residual plot, we can see by eye that this calibration does not hold well everywhere. As we will describe in the following section, we performed multiple colour and magnitude cuts to determine a selection for which the sample of KOI host stars is representative of the larger KIC population, as well as such that the metallicity callibration is well behaved. We determined that this range is:
\begin{align*}
    1.4 \leq &u-b \leq 2.8\\
    1.2 \leq &G-K \leq 2.0
\end{align*}
The residuals for the colour cut are shown in blue in Fig. \ref{img_callibration}, with a smaller standard deviation of 0.16\,dex.

\begin{figure*}
  \centering
  \includegraphics[width=\textwidth]{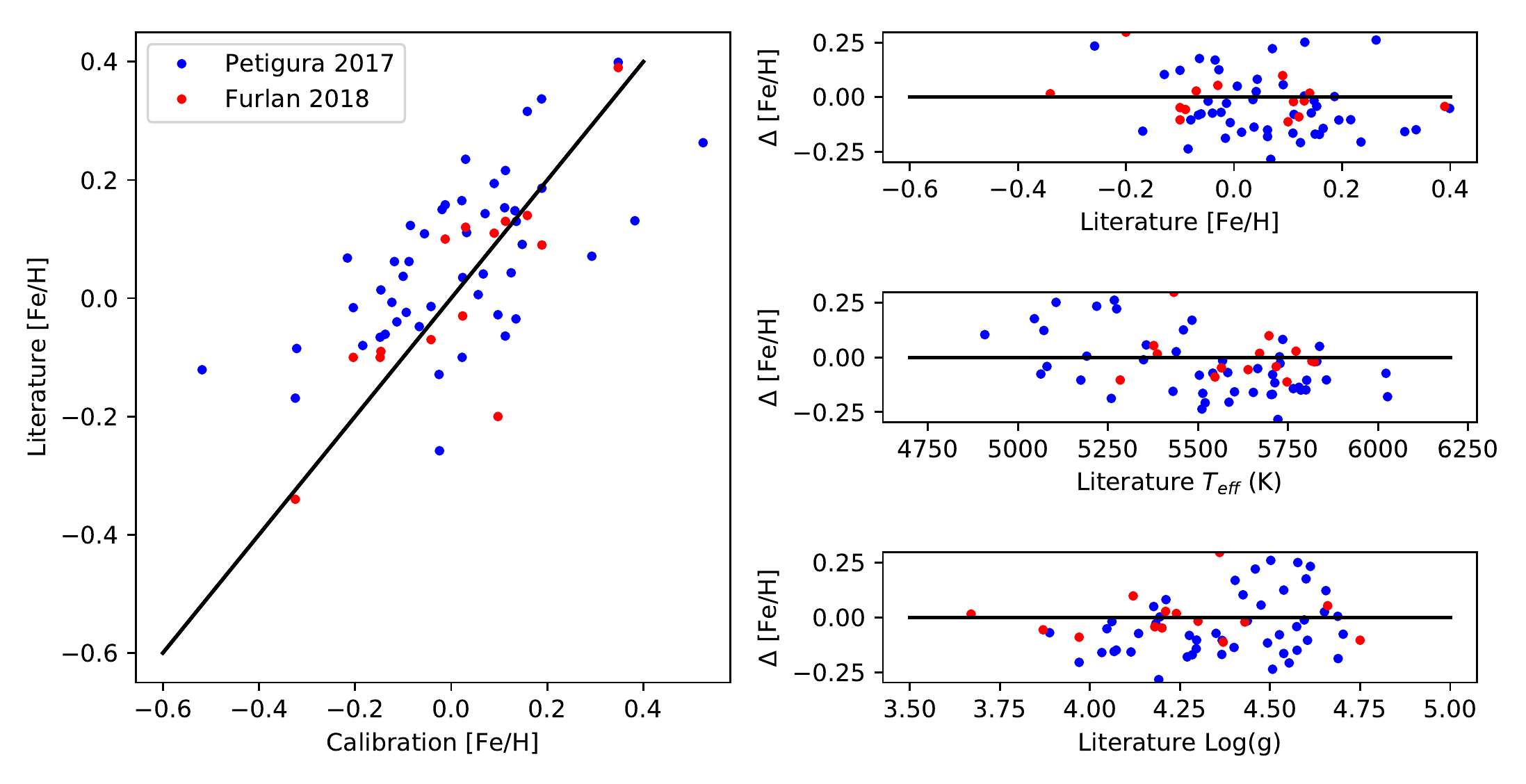}
  \caption{Comparison of our photometric metallicity against that of \protect\cite{petigura} and \protect\cite{furlan_kepler_2018}, with residuals plotted against the literature values of metallicity, effective temperature and surface gravity.}
  \label{img_petigura}
\end{figure*}

To test the validity of the calibration, our photometric metallicities were tested against the spectroscopic metallicities measured by \cite{petigura} and \cite{furlan_kepler_2018}. It should be noted that these samples comprise of mostly main-sequence stars; the regime where most planet host stars reside. The comparisons are shown in Fig. \ref{img_petigura}, with a mean difference (Our metallicity - \cite{petigura}) of $-0.05 \pm 0.14$\,dex, and (Our metallicity - \cite{furlan_kepler_2018}) of $0.00 \pm 0.10$\,dex. Both of these are well within the quoted uncertainty of our metallicity calibration. \red{There is an indication that the scatter increases towards cooler $T_{\rm{eff}}$ and higher $\log(g)$. As spectroscopic parameters are harder to determine for cooler stars, the increased scatter could be due to deficiencies in our calibration sample, as well as in the spectroscopic samples we compare against.}

\section{Determining a Representative Sample}
\label{sec:representative}

\red{When using a sample to perform population studies, it is important to assess how well such a sample is representative of the underlying population of stars in the field. In this case, the underlying population of stars in the \textit{Kepler} field is that assembled through the KIC catalogue, whereas our population inferences are derived using the KOI sample (note that in both cases a cross-match against \textit{Gaia} and Str\"omgren photometry is required, see discussion in Section \ref{sec:catalogue}).} For e.g., in comparing the metallicity distribution of the KOI sample to the KIC sample, it is important to understand the extend of any differences in brightness or colour distributions \red{between the two samples, as this could bias conclusions} about planetary demographics. \red{If the KIC sample were to be extend to fainter magnitudes, it would trace stars further away in the Galaxy, and differences in metallicity with respect to the KOI would also stem from Galactic metallicity gradients \citep[e.g.,][]{boeche13,mikol}. Likewise, if the KIC sample were to extend to bluer colours, it would encompass early-type, young stars whose metallicities span a limited range, reflecting the chemistry of the present day ISM \citep[e.g.,][]{nieva,luck}}. 
%These differences could be potentially caused by a strong stellar-mass dependent planetary frequency \citep[e.g., as is know for rare Jovian planets][]{bowler_retired_2010}, or a strong effect of instrumental noise in planet detectability at fainter included KIC stars. 
If however the KIC and KOI samples are very similar in apparent magnitude and colour distributions, we can perform a meaningful comparison between their metallicities.

Since our sample is drawn from photometric catalogues, we can perform well defined magnitude and colour cuts and ensure the KOI sample represents the underlying sample of stars found in the KIC catalogue. \red{This is different from spectroscopically selected samples, where stars are picked for their KOI status at the time observations are done, and often favouring brighter targets. For example, in the California-Kepler Survey \citep{petigura}, spectroscopy was obtained for KOI brighter than $14.2$, with fainter targets appended for a variety of reasons. In contrast, with a photometric sample is straightforward to observe stars to a given magnitude limit, regardless of how the \textit{Kepler} planet catalogue grows in size with time}. 

\begin{figure}
  \centering
  \includegraphics[width=\linewidth]{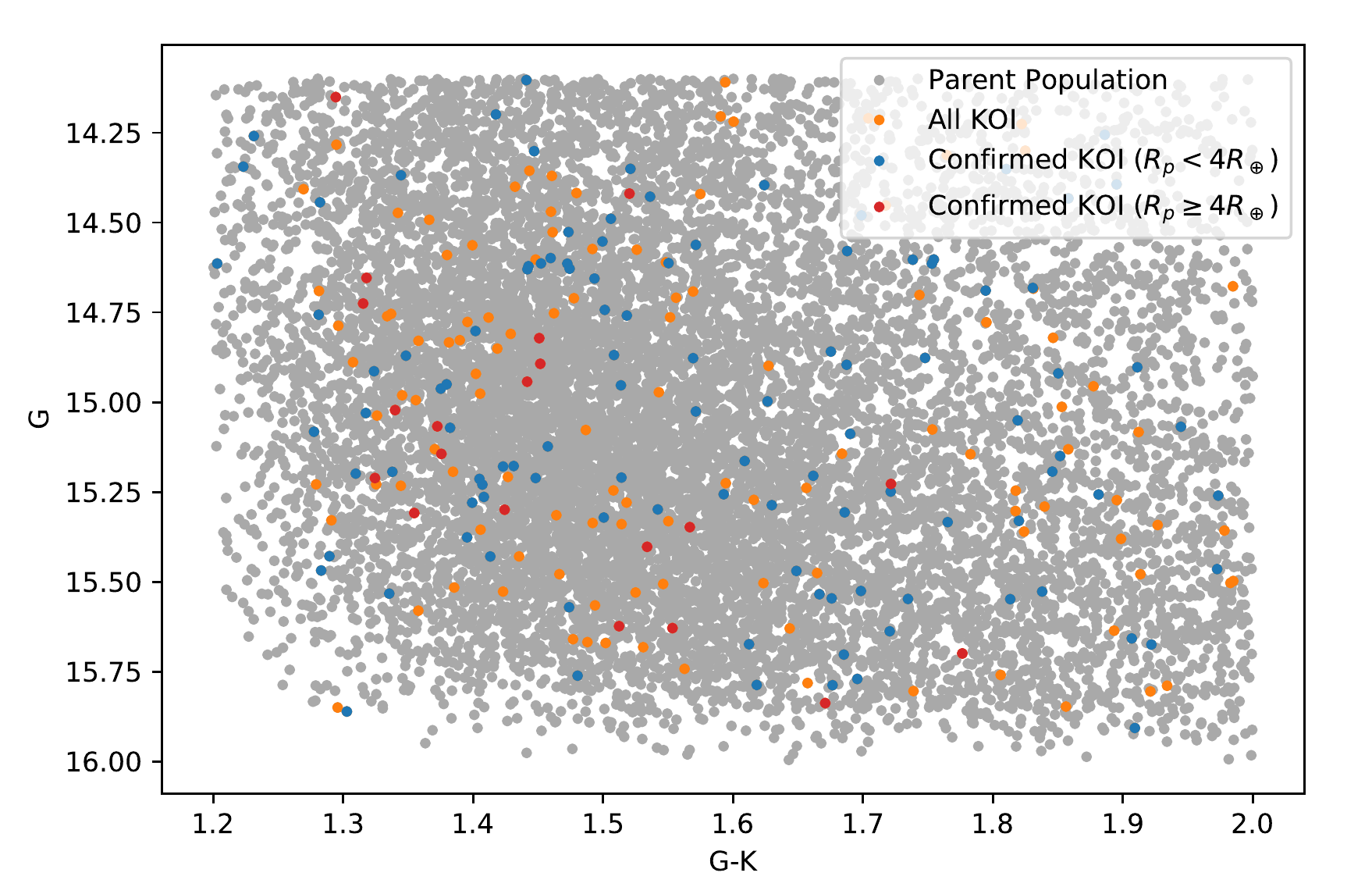}
  \caption{Comparison between the full photometric catalog and KOI (all and confirmed) observed over the same sky footprint.}
  \label{img_HR}
\end{figure}

\red{To derive colour and magnitude cuts for the KIC and KOI samples, we applied the methodology described in \cite{c16}, who used it to build an unbiased sample of asteroseismic targets in the \textit{Kepler} field.
We started by taking the full sample of KOI, and benchmark their distribution in $(G-K)$ and $(u-b)$ colours, as well as apparent $G$ magnitudes against the KIC sample in the same ranges. To this purpose we used the Kolmogorov-Smirnov (KS) statistic on the colour and magnitude distributions. The significance levels between the KOI and the KIC sample pass from being virtually zero when stars are selected regardless of their colours and magnitudes, to order 20 to 70 percent when using the cuts listed below:
\begin{align*}
    1.4 \leq &u-b \leq 2.8\\
    1.2 \leq &G-K \leq 2.0\\
    14.1 \leq &G \leq 16.0
\end{align*}
This implies that the null assumption that the two samples are drawn from the same population cannot be rejected.
These cuts were determined by exploring different ranges and each time running a KS test. We checked that similarly high percentages were also obtained if using a different diagnostic such as the Wilcoxon rank-sum test. Other colour combinations were also explored, although we focus our discussion on $(G-K)$ and $(u-b)$ since these indices underpin our metallicity calibration. The results are summarised in Table \ref{Table_sample_all}, and were repeated for the subset of KOIs that have a \textsc{confirmed} disposition according to the NASA Exoplanet database (Table \ref{Table_sample_conf})}. In the following of the paper the full photometric catalogue will be referred to as the {\it parent population} when restricted to the above colour and magnitude ranges, and its comparison aginst the KOI and confirmed KOI is shown in Fig. \ref{img_HR}.

\begin{table}
\centering
\caption{KS statistic, Wilcoxon rank-sum statistic and resulting p-values of the distribution of colours and magnitudes between all KOIs and the full photometric sample when the cuts discussed in Section \ref{sec:representative} are applied.}
\label{Table_sample_all}
\resizebox{\linewidth}{!}{%
\begin{tabular}{ccccc}
\hline
\textbf{Parameter} & \textbf{KS Statistic $D$} & \textbf{KS $p$} & \textbf{Wilcoxon Statistic $U$} &\textbf{Wilcoxon $p$} \\ \hline
$G-K$          & 0.047     & 0.667   & 0.408 & 0.683     \\
$u-b$          & 0.053     & 0.499   & 0.798 & 0.424          \\
$G$            & 0.043     & 0.756   & 0.217 & 0.828         \\ \hline
\end{tabular}%
}
\end{table}
\begin{table}
\centering
\caption{KS statistic, Wilcoxon rank-sum statistic and resulting p-values of the distribution of colours and magnitudes between all confirmed KOIs and the full photometric sample when the cuts discussed in Section \ref{sec:representative} are applied.}
\label{Table_sample_conf}
\resizebox{\linewidth}{!}{%
\begin{tabular}{ccccc}
\hline
\textbf{Parameter} & \textbf{KS Statistic $D$} & \textbf{KS $p$} & \textbf{Wilcoxon Statistic $U$} &\textbf{Wilcoxon $p$} \\ \hline
$G-K$          & 0.061    &  0.714    & 0.233 & 0.671     \\
$u-b$          & 0.091     & 0.233   & -1.575 & 0.115          \\
$G$            & 0.062     & 0.709    & 0.425 & 0.816         \\ \hline
\end{tabular}%
}
\end{table}

\section{Obtaining Radii and Masses}
\label{sec:radii}
One trend we aimed to investigate was the ``Planet-Radius gap'', a feature where there is a relative absence of planets with radii around 1.9~$R_{\oplus}$ \citep[e.g.,][]{owen_kepler_2013,fulton_california-kepler_2017}, with some studies showing that the depression follows a slight dependence on the mass of the host star \citep[e.g.,][]{ fulton_california-kepler_2018,berger_gaia-kepler_2020-2}.

We derived stellar masses and radii using the Bayesian isochrone fitting algorithm \textsc{Elli} \citep{lin_stellar_2018}, which is built upon the MIST isochrones \citep{2016ApJ...823..102C}. The input parameters used by \textsc{Elli} are effective temperatures ($T_{\rm{eff}}$), 2MASS $K$ magnitudes, reddening, \textit{Gaia} DR2 parallaxes, surface gravities $\log(g)$ and our photometric metallicities. In the following, we describe in detail our procedure.

To obtain effective temperatures we run the InfraRed Flux 
Method (IRFM) for all our KOIs. The IRFM is an almost model independent photometric technique originally devised to obtain angular diameters to a precision of a few per cent, and capable of competing against intensity interferometry should a good flux calibration be achieved \citep[e.g.,][]{black77,black80}. We 
used the implementation described in \cite{c20} which employs \textit{Gaia} and 2MASS photometry to derive effective temperatures and angular diameters for stars of known metallicity and surface gravity. We adopted our photometric metallicities, and $\log(g)$ from the KOI catalogue. Effective temperatures derived from the IRFM were then fed into \textsc{Elli} along the other parameters needed to derive stellar radii and masses. A new estimate of $\log(g)$ was computed, iterating between the IRFM and \textsc{Elli}. Because of the mild dependence of the IRFM on the adopted metallicity and surface gravity \citep[see e.g.,][]{Alonso95,c06} only a couple of iterations were necessary to converge on a final mass and radius for each star.

\begin{figure}
  \centering
  \includegraphics[width=\linewidth]{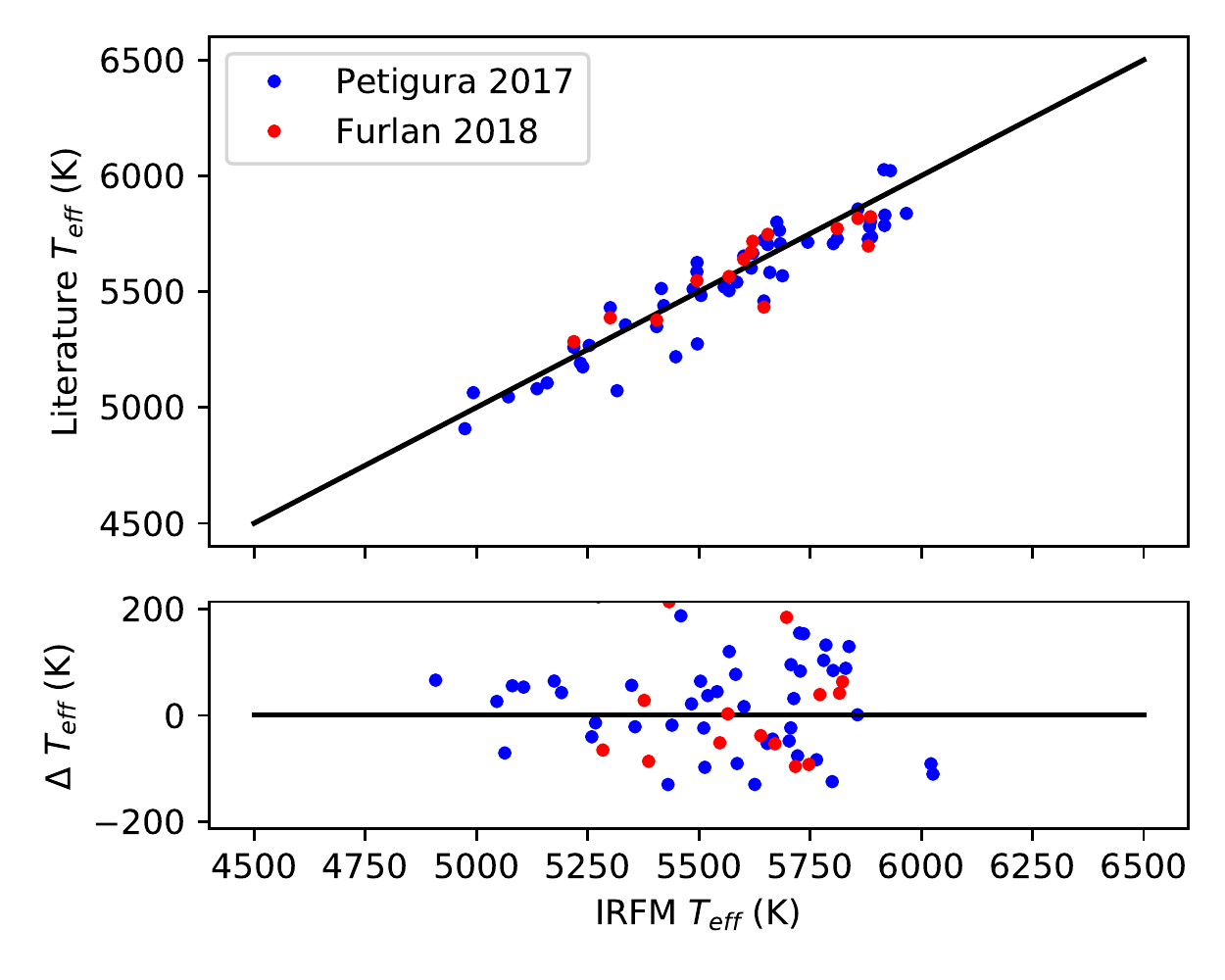}
  \caption{Comparison of our effective temperatures from the IRFM (abscissa) with those of \protect\cite{petigura} and \protect\cite{furlan_kepler_2018}.}
  \label{img_furlan_teff}
\end{figure}

First, we compared our $T_{\rm{eff}}$ against those published in 
\cite{petigura} and \cite{furlan_kepler_2018}, showing excellent agreement, with a mean difference of 30\,K and a standard deviation of 90\,K (Fig. \ref{img_furlan_teff}). Since $T_{\rm{eff}}$ from the IRFM are sensitive to reddening (where a change of $\pm 0.01$ in $E(B-V)$ has an impact of $\pm 50$~K), this comparison suggests that reddening is well under control. 

In addition to stellar radii obtained from isochrone fitting, 
the availability of angular diameters from the IRFM and \textit{Gaia} distances \citep{Bailer-Jones2018} for all our targets allowed us to derive radii independently of stellar isochrones. We dub these "empirical radii" since they are virtually free from any stellar modelling assumption. \red{Since distance uncertainties propagate into radii, from now on we apply a very mild cut on parallax uncertainty to remove stars with clearly ill measured parallaxes. At the same time, we do not want to apply any stringent parallax requirement, as this could potentially introduce an extra sample selection effect. We only require stars to have parallaxes better than 20 percent, but in fact the vast majority of our targets have distances determined at a few percent level}. Finally, the planet radius was determined by applying the planet to star radius ratio provided in the KOI catalogue; a parameter estimated from the transit depth.

\begin{figure}
  \centering
  \includegraphics[width=\linewidth]{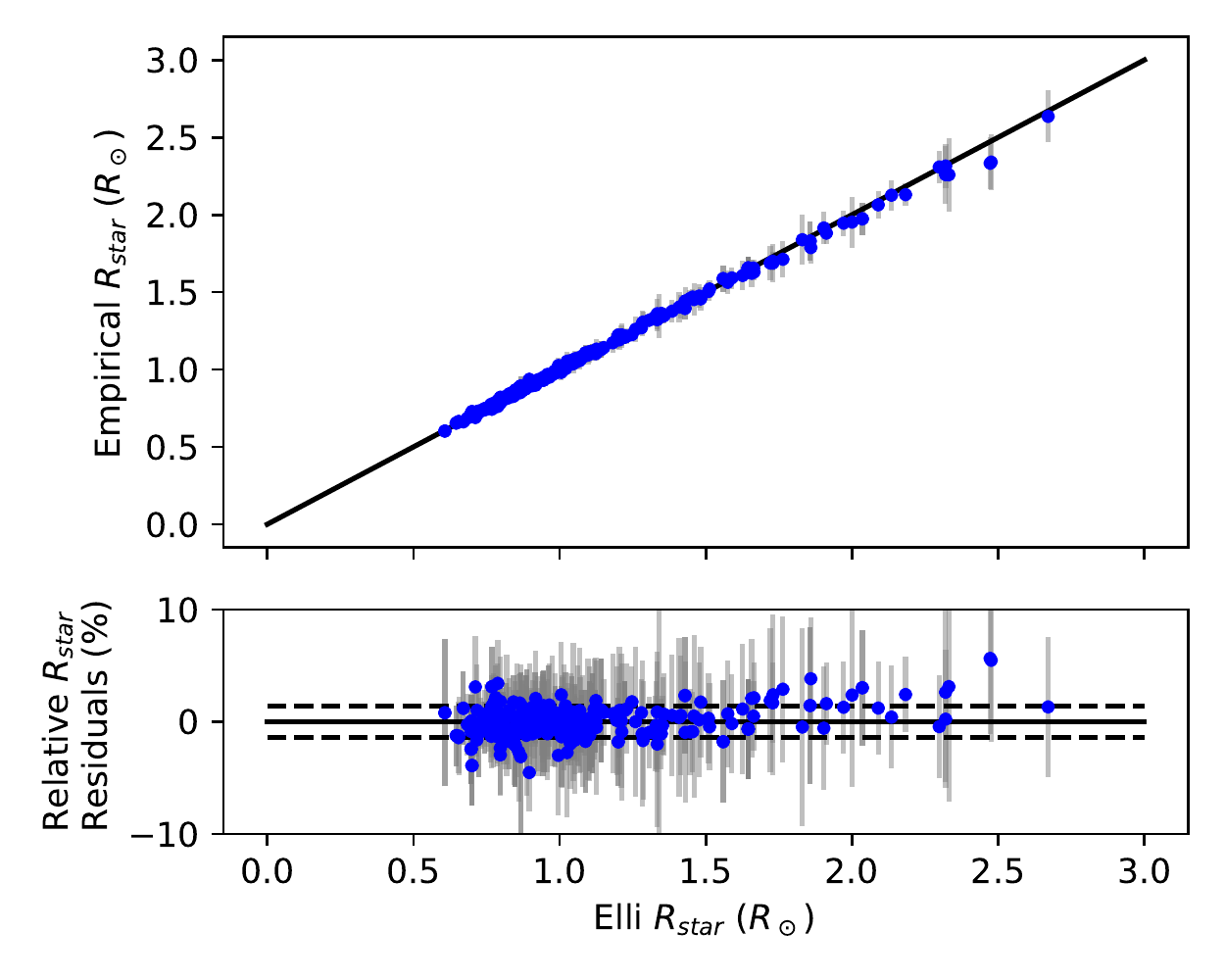}
  \caption{Comparison of the stellar radii derived from \textsc{Elli} against the empirical ones obtained using the \protect\cite{Bailer-Jones2018} distances and angular diameters from the IRFM. \red{Uncertainties are shown in grey, with dashed lines indicating 1$\sigma$ scatter.}}
  \label{img_cbj_radius}
\end{figure}

Fig. \ref{img_cbj_radius} shows the relative difference between the stellar radii derived from \textsc{Elli} and the empirical ones
--(\textsc{Elli} radius - empirical radius)/\textsc{Elli} radius-- with a a mean of $0.00\pm 0.03$ solar radii, and a \red{standard deviation} of 2\,percent. This gives confidence that the radii and other stellar quantities derived from \textsc{Elli} are robust \red{and that possible systematic differences arising from our methodology are within our quoted uncertainties}. From this point on, we adopt the empirical radii as our accepted stellar radii.

The empirical radii also have good relative uncertainties, with a mean of 3.4\,percent, which is on par with that of \cite{berger_gaia-kepler_2020-1} and \cite{fulton_california-kepler_2018}. When multiplied by the \textit{Kepler} planet to star radius ratio, we find our planet radii have uncertainties with a mean of 6.2\,percent, highlighting that uncertainties in the \textit{Kepler} radius ratio carry a significant contribution to the uncertainties in the planetary radius.

\begin{figure}
  \centering
  \includegraphics[width=\linewidth]{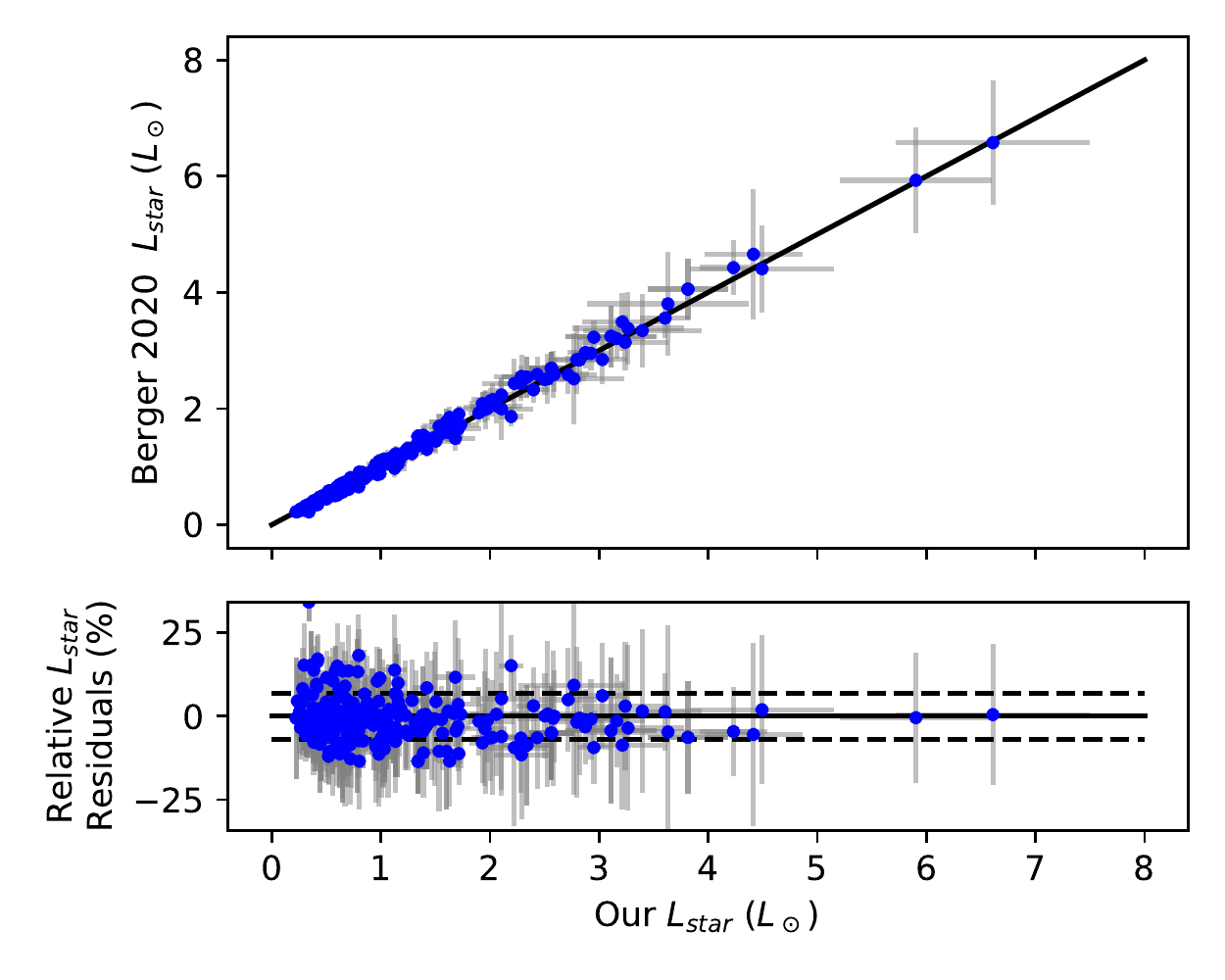}
  \caption{Comparison of our stellar luminosities with those of \protect\cite{berger_gaia-kepler_2020-1}. \red{Uncertainties are shown in grey, with dashed lines indicating 1$\sigma$ scatter.}}
  \label{img_berger_lum}
\end{figure}

\begin{figure}
  \centering
  \includegraphics[width=\linewidth]{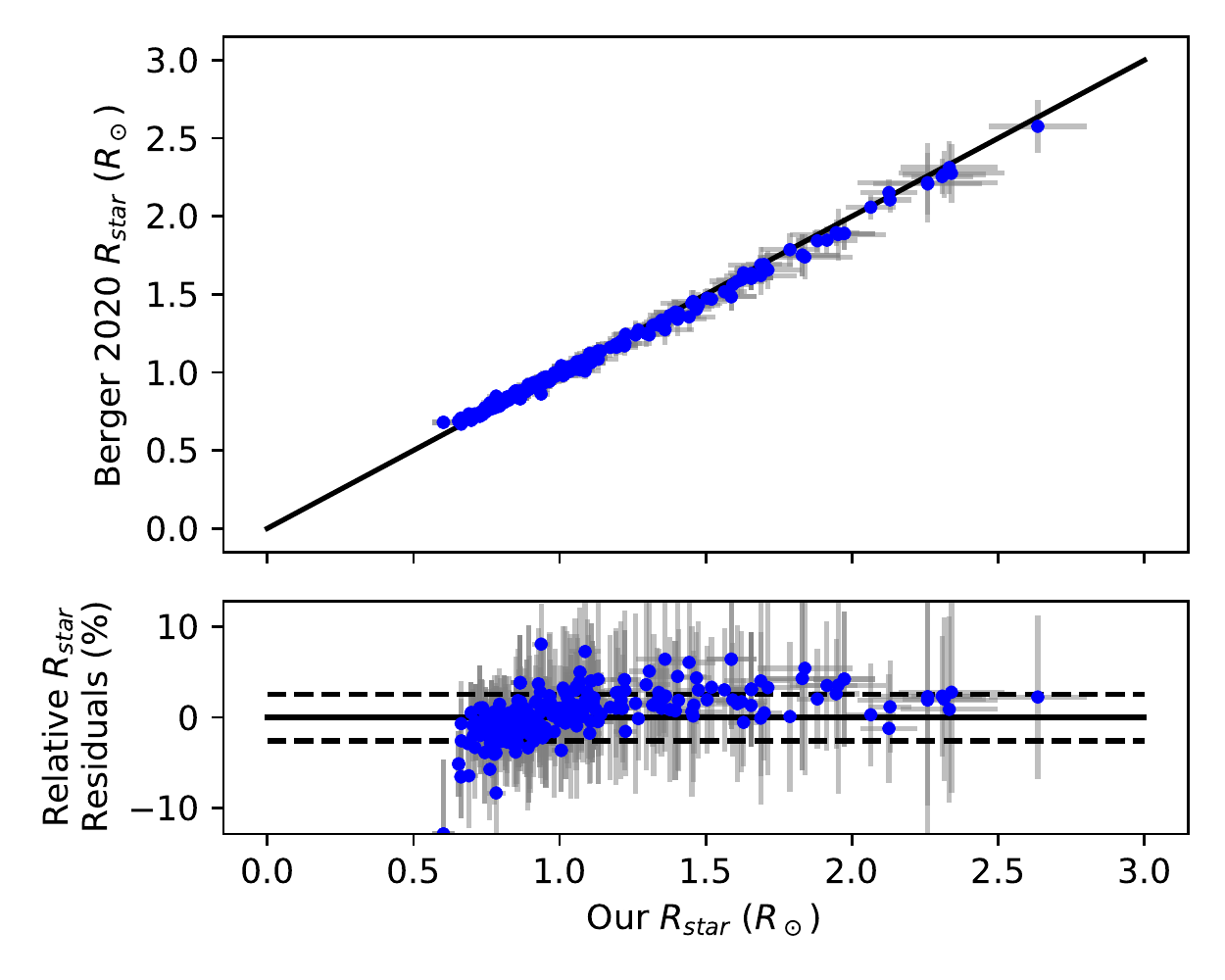}
  \caption{Comparison of our stellar radii with those of \protect\cite{berger_gaia-kepler_2020-1}. \red{Uncertainties are shown in grey, with dashed lines indicating 1$\sigma$ scatter.}}
  \label{img_berger_srad}
\end{figure}

Finally, we tested our stellar parameters against those derived by \cite{berger_gaia-kepler_2020-1}, in particular the stellar luminosity (Fig. \ref{img_berger_lum}) and stellar radius (Fig. \ref{img_berger_srad}). Both of these parameters had extremely good agreement, with luminosity residuals of \red{$-1\pm 7~\%$} and radius residuals of \red{$0.4\pm 2.6~\%$; any trends within these residuals are less than the order of these uncertainties.} We also tested our masses against their catalogue, finding a mean difference and standard deviation of $-6 \pm 7$~percent, which again shows good agreement albeit with our masses tending to be lower than those of \cite{berger_gaia-kepler_2020-1}.

\section{Results and Analysis}
\label{sec:results}
\subsection{Metallicity trends}
\label{sec:metallicity_results}
\begin{figure*}
  \centering
  \includegraphics[width=\textwidth]{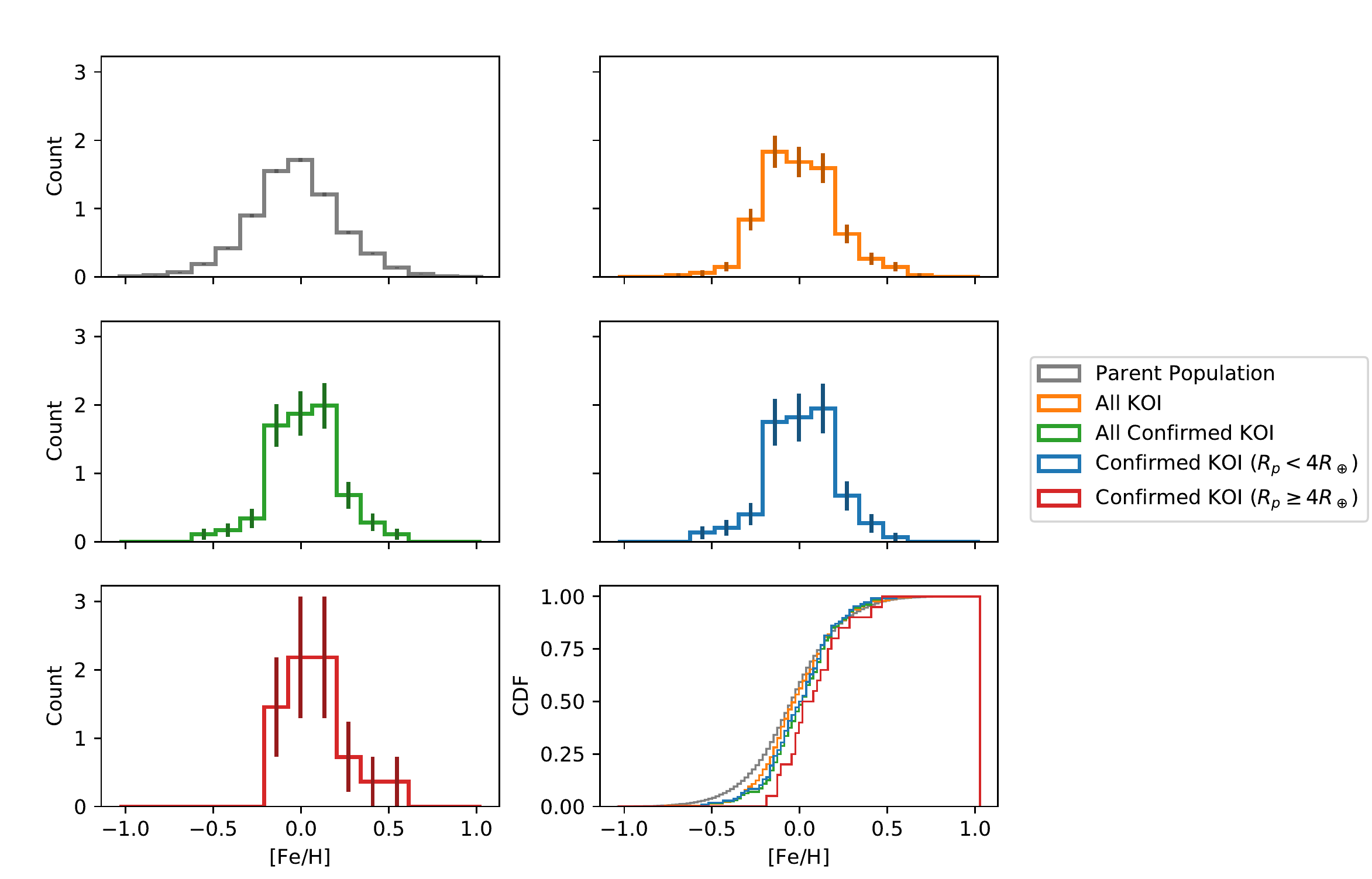}
  \caption{Histograms and cumulative distribution functions of the metallicity distribution of various exoplanet subsets. \red{Histograms are normalised with an area of one.} Confirmed KOI are those with a NASA Exoplanet Database designated disposition of \textsc{confirmed}, whereas ``All KOI'' refer to those with a disposition of \textsc{confirmed} or \textsc{candidate}.}
  \label{img_fe_histcdf}
\end{figure*}

We first investigated trends concerning metallicity between the KOI host stars and the {\it parent population} (ref. Section \ref{sec:representative}). The subset of KOIs which have been labelled as \textsc{confirmed} by the NASA Exoplanet Database was also extracted and compared, before finally splitting the confirmed KOIs between large ($R_p \geq 4 R_\oplus$) and small ($R_p < 4R_\oplus$) planetary radii. The use of the confirmed sub-sample was to ensure that we were not affected by non-planetary companions, with the trade off of a smaller sample size. If a star had multiple planets, then it was classified according to the radius of the largest one. \red{We note here that our sample mainly consists of sub Neptunes and super Earths with radii between 1 and 10 $R_\oplus$, along with a few larger and smaller exoplanets}. Histograms and cumulative distribution functions (CDFs) of these populations are shown in Fig. \ref{img_fe_histcdf}.

The mean metallicity of the KOIs ([Fe/H] = -0.01) and especially that of the confirmed KOIs ([Fe/H] = 0.02) is different from that of the {\it parent population} of stars as a whole ([Fe/H] = -0.03). This suggests that the exoplanet host stars are more metal rich than the rest of the candidate KOIs. To confirm this deviation, KS and Wilcoxon rank-sum tests were undertaken using the full metallicity distribution function (MDF), with each subset being tested against the parent population. The results are shown in Table \ref{Table_fe}.

\begin{table}
\centering
\caption{KS tests of the metallicities of the different subsets against the metallicies of the full photometric sample in the same colour and magnitude ranges.}
\label{Table_fe}
\resizebox{\linewidth}{!}{%
\begin{tabular}{ccccc}
\hline
\textbf{Sample} & \textbf{KS Statistic $D$} & \textbf{KS $p$} & \textbf{Wilcoxon Statistic $U$} &\textbf{Wilcoxon $p$} \\ \hline
All KOI host stars               & 0.081     & 0.087   & -1.629 & 0.103     \\
All confirmed KOI host stars               & 0.155     & 0.004   & -2.879 & 0.004     \\
Confirmed KOI ($R_p < 4R_\oplus$)    & 0.141     & 0.025   & -2.112 & 0.035          \\
Confirmed KOI ($R_p \geq 4R_\oplus$) & 0.308     & 0.035   & -2.389 & 0.017         \\ \hline
\end{tabular}%
}
\end{table}

With a $p \simeq$ 9~percent, the null hypothesis that the MDF of all KOI is drawn from that of the {\it parent population} cannot be rejected. However, when restricting ourselves to the sample of confirmed KOIs, the \red{$p$ value} drops to a mere $0.4$~percent, thus allowing us to reject the null hypothesis \red{at a very high significance}. This is also the case for the two sub-samples with small and large planetary radii, \red{where we can reject the null hypothesis at the 5\% significance level}. In particular, when looking at the histograms in Fig. \ref{img_fe_histcdf}, we see that the confirmed exoplanet host stars seem to favour higher metallicities than their non-exoplanet hosting counterparts, supporting earlier results \citep[e.g.,][]{gonzalez97,santos_spectroscopic_2004,fischer_planet-metallicity_2005,zhu_influence_2019}. The larger p-value from the sample of all KOIs (including those with a disposition of \textsc{candidate}), may be due to some of the candidate KOIs not being planetary companions. 

\begin{figure}
  \centering
  \includegraphics[width=\linewidth]{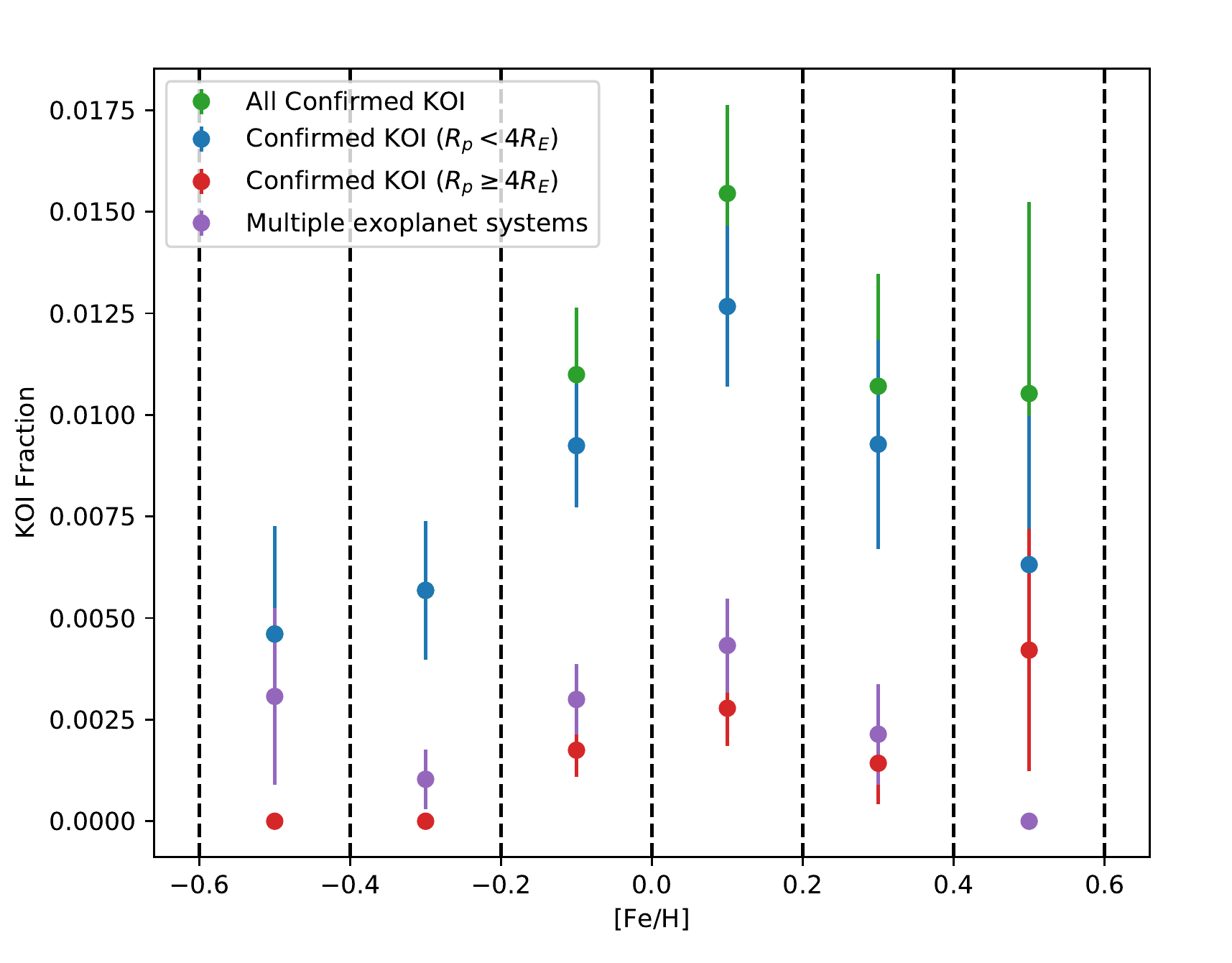}
  \caption{Fraction of the parent population that host exoplanets for a given metallicity bin. Different colours refer to the different planet sizes of the KOI, or whether they are present in a system with multiple objects. Error bars are drawn from the Poisson statistic.}
  \label{img_paper_plot}
\end{figure}

To investigate further the trend with metallicity, a plot of the percentage of stars with a confirmed KOI for a given metallicity bin was created. This particular diagram was based on the work of \cite{zhu_influence_2019}, and can be seen in Fig. \ref{img_paper_plot}. Also plotted in this figure is the percentage of stellar systems with multiple exoplanets, with the aim to compare to the findings by \cite{brewer_compact_2018} as to whether multiple exoplanet systems were preferentially around lower metallicity host stars. 

As predicted from \cite{santos_spectroscopic_2004}, \cite{fischer_planet-metallicity_2005} and \cite{zhu_influence_2019} among others, and as inferred from the cumulative distribution function, exoplanets (and especially those that have a radius greater than 4 Earth radii) appear to form preferentially around higher metallicity stars. Smaller exoplanets also appear to favour higher metallicities, peaking above solar metallicity, and then declining at very high metallicities \red{although in our case the significance of this trend is only at 1 $\sigma$ level and so it should be considered carefully}. This furthers the work of \cite{buchhave_abundance_2012}, who claims that while smaller exoplanets form around stars with a wide range of metallicities, large exoplanets form around primarily metal rich stars. We show that while the smaller exoplanets have a weaker trend than their larger counterparts, they still have a bias towards metallicities around and above solar metallicity. 

We summarise these findings in Table \ref{Table_zhu}, which shows the percentage of stars in our representative sample of the \textit{Kepler} field that host exoplanets for metallicities above and below solar metallicity. Here, \red{to almost 2$\sigma$ significance} we find that large exoplanets are more than twice as likely to be found around metal rich stars while smaller exoplanets are 1.5 times as likely. We thus find that all exoplanets are more likely to be observed at higher metallicities, with the size of the exoplanet influencing the strength of this trend. 

\red{Multiple planet systems also have a peak at solar metallicity. The upward trend in the lowest metallicity bin while intriguing has a too large uncertainty to allow any meaningful comparison with \cite{brewer_compact_2018}, who found that compact multi-planet systems occur more frequently around stars of increasingly lower metallicities.}

\begin{table}
\centering
\caption{Percentage of stars in our representative sample of the \textit{Kepler} field hosting KOIs for high ([Fe/H] $\geq$ 0) and low ([Fe/H] < 0) metallicities. Uncertainties are drawn from the Poisson statistic.}
\label{Table_zhu}
\resizebox{\linewidth}{!}{%
\begin{tabular}{ccc}
\hline
\textbf{Sample} & \textbf{Metal Poor [Fe/H] < 0 (\%)} & \textbf{Metal Rich [Fe/H] > 0 (\%)}  \\ \hline
All confirmed KOI host stars     & 0.88 $\pm$ 0.12     &  1.37 $\pm$ 0.16   \\
Confirmed KOI ($R_p < 4R_\oplus$)    & 0.77 $\pm$ 0.11     & 1.12 $\pm$ 0.15          \\
Confirmed KOI ($R_p \geq 4R_\oplus$) & 0.11 $\pm$ 0.04     & 0.25 $\pm$ 0.07         \\
Multiple exoplanet systems      & 0.24 $\pm$ 0.06 & 0.33 $\pm$ 0.08 \\\hline
\end{tabular}%
}
\end{table}

\begin{figure*}
  \centering
  \includegraphics[width=\linewidth]{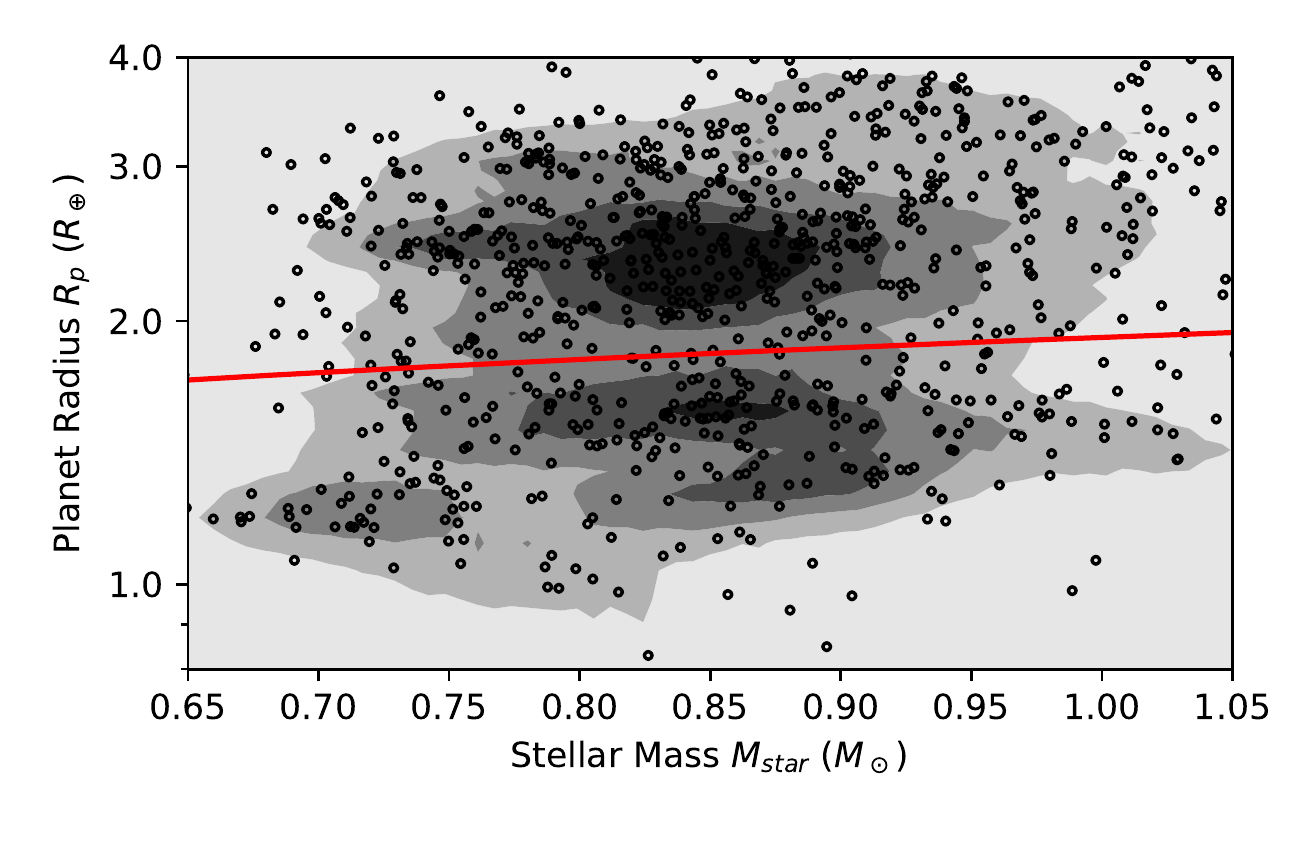}
  \caption{2D Monte Carlo distribution of the stellar mass and planetary radius of KOI with a \textsc{confirmed} disposition. Density contours are shown behind a random selection of 1000 samples. Note the presence of a gap around $R_p = 1.9\,R_\oplus$. Also plotted in red is the best fit line to the radius gap from \protect\cite{berger_gaia-kepler_2020-2}, with their slope of $d\log{R_p}/d\log{M_{star}} = 0.26$. It can be seen that our data also fits this line well, showing that the radius gap has a slight positive correlation with stellar mass.  }
  \label{img_planet_gap}
\end{figure*}

\subsection{The planet-radius gap}
\label{sec:radius_results}
As mentioned previously, one aim of this work was to study the planet-radius gap. With the stellar mass and planet radius we recovered through the processes outlined in Section \ref{sec:radii}, we generated a 2D density plot through a Monte Carlo (MC) simulation. We drew 10,000 samples assigning each time normal errors in stellar mass and planet radius for each of our confirmed KOI data points, and plotted the density distribution as contours in Fig. \ref{img_planet_gap}. We also plot 1000 random samples from the MC simulation. We chose not to include the KOIs with a \textsc{candidate} disposition due to the potential presence of false positives in this sample (examined in Section \ref{sec:metallicity_results}). 

\red{The contour levels suggest the presence of a gap around 1.8-2.0~$R_\odot$, with a mild positive slope as function of stellar mass. This trend has been found in previous works by \cite{fulton_california-kepler_2018} and \cite{berger_gaia-kepler_2020-2}. Indeed, if in Fig. \ref{img_planet_gap} we overplot the best fit line to the radius gap from \cite{berger_gaia-kepler_2020-2} (slope $d\log{R_p}/d\log{M_{star}} = 0.26$), this line also cuts across our gap.}

\begin{figure}
  \centering
  \includegraphics[width=\linewidth]{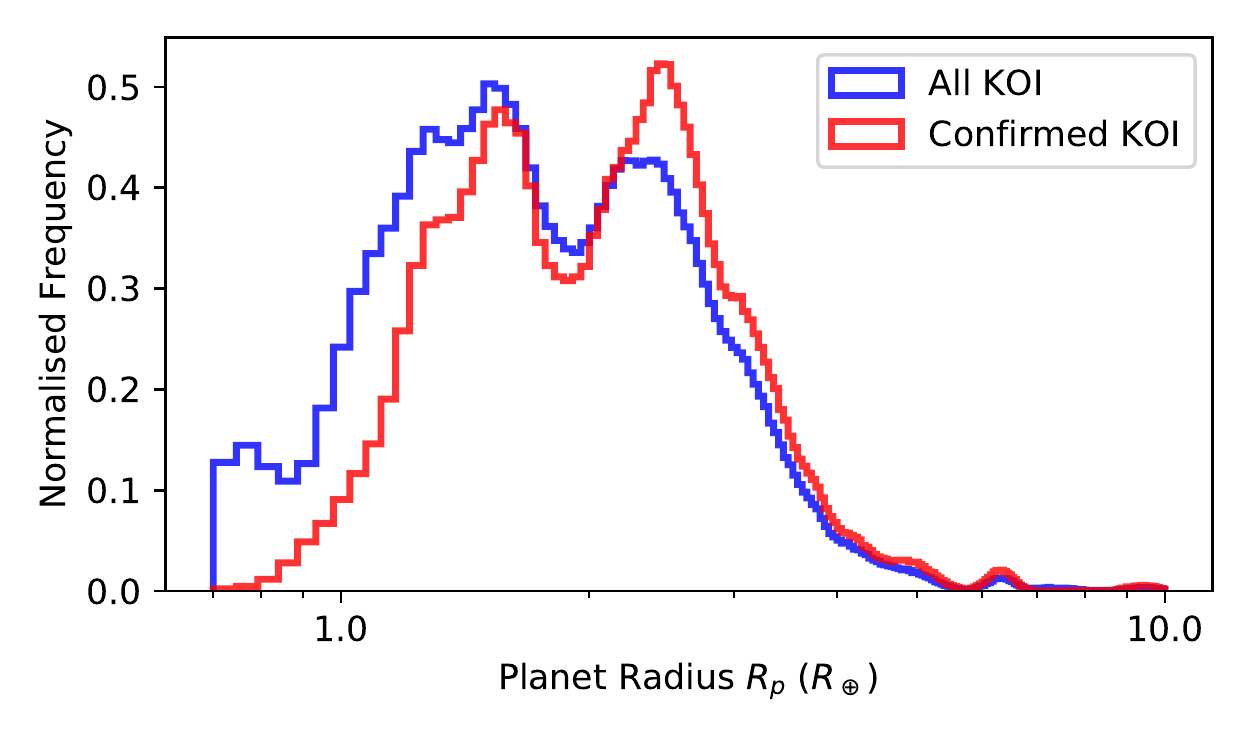}
  \caption{Normalised histograms of the planetary radius of all KOI (blue) and only the confirmed KOI (red).}
  \label{img_debias_hist}
\end{figure}

To view the gap more clearly, we contracted this plot over mass and normalised the data, creating a simple histogram of planet radii. This is shown in solid red in Fig. \ref{img_debias_hist}. We also show in solid blue the histogram of planet radii when KOI with a disposition of \textsc{candidate} are included. The most notable feature is a very clear bimodal distribution, with a gap again at 1.9~$R_\oplus$, supporting the conclusions of e.g \cite{fulton_california-kepler_2018} and  \cite{berger_gaia-kepler_2020-2}. The restriction of only including confirmed KOI influences the distribution by increasing the side of the second peak at $\sim2.5~R_\oplus$ and decreasing the width of the first at $\sim1.6~R_\oplus$, but the location of the gap does not change. 

\begin{figure}
  \centering
  \includegraphics[width=\linewidth]{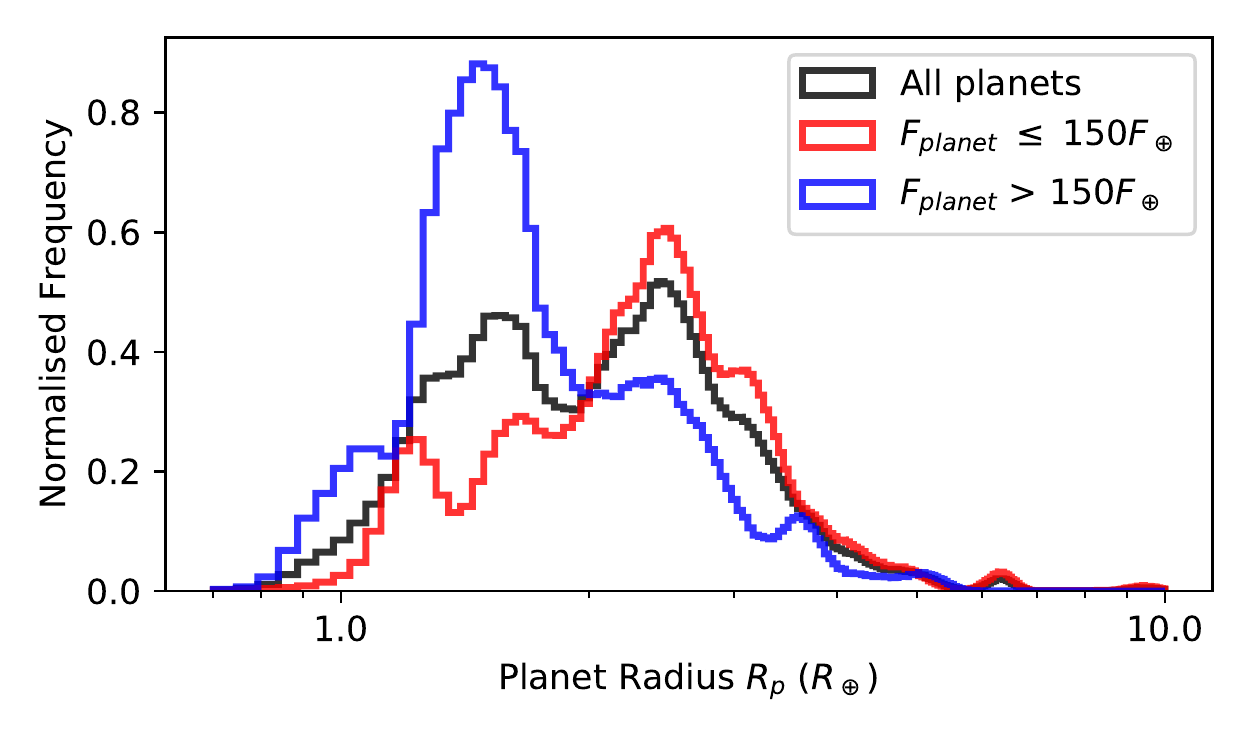}
  \caption{Normalised histograms of the planetary radius of confirmed KOI, separated by host star luminosities. All confirmed KOI are shown in black, whereas the red and blue histograms represent the subset with a planetary flux of less than and greater than 150\,$F_\oplus$ respectively.}
  \label{img_planet_lum_hist}
\end{figure}

Following the work of \cite{berger_gaia-kepler_2020-2}, we also looked at how the radius histogram was affected by the incident stellar flux falling on the planet. We chose the separating flux value of 150\,$F_\oplus$ from \cite{berger_gaia-kepler_2020-2}, to test to see whether we identified a similar trend; our sample had 114 planets designated as cool. This is plotted in Fig. \ref{img_planet_lum_hist}, where again we have chosen to only plot confirmed KOI. We recover that planets with higher incident flux exhibit smaller radii than their cooler counterparts, possibly due to evaporation of the atmospheres of these planets. As with \cite{berger_gaia-kepler_2020-2}, we caution that these results may be a result of small number statistics and are likely fraught with \textit{Kepler} selection effects. 

\section{Conclusions}

We have compiled a photometric catalogue of stars in the \textit{Kepler} field utilising photometry from \textit{Gaia}, Str\"omgren and 2MASS catalogues. We created a metallicity calibration based on APOGEE spectroscopy to obtain a metallicity for our photometric sample, and then performed well defined colour and magnitude cuts to ensure our dataset was a representative sample of the underlying population of stars. We then derived temperatures and angular diameters using the IFRM, which were then used to derive stellar radii and stellar mass through Bayesian isochrone fitting. The resultant parameters were compared favourably with previous results from the literature, especially giving stellar radii with relative uncertainties around 3.4~percent. Planetary radii uncertainties of 6.2~percent hence indicate a major uncertainty contribution from the \textit{Kepler} planet to star ratio.

\red{The main purpose of this work has been to outline a methodology to derive photometric stellar parameters for planet host stars. The advantage of using photometry is that it allows a better control of sample selection effect, and in the future we believe to more robust inferences on population studies.} Although our sample is currently relatively small, especially compared to recent studies such as that of \cite{fulton_california-kepler_2018} and \cite{berger_gaia-kepler_2020-2}, we are able to recover a number of known trends. 

We find that the stars hosting confirmed KOIs have a statistically different metallicity distribution than the parent population of stars in the \textit{Kepler} field. We also find that this statistical claim is not valid for the sample of KOIs that include those with a disposition of \textsc{candidate}, consequences of \red{potential} undetected false positives in the list of KOIs. 

We quantify the metallicity distribution differences between KOI and the larger sample of KIC stars, finding that KOI hosts tend to be more metal rich than their non-planet hosting counterparts. While holding especially true for large exoplanets larger than $4~R_\oplus$, which has been known about in literature for some time \citep[e.g,][]{buchhave_abundance_2012}, we also find this holds for smaller exoplanets albeit to a weaker extent. This follows the conclusions of \citep[e.g,][]{zhu_influence_2019}. 
    %Finally, we also find that \textit{Kepler} stars with more than one exoplanet favour metal rich planets to a lesser degree than those with a single detection, supporting the conclusions of \cite{brewer_compact_2018}. In particular, we find an increase in the number of multiple exoplanet systems at the lowest metallicity bin in our sample.

We recover the planet radius gap at $\sim 1.9~R_\oplus$. We also tentatively find that there is a trend that planets with a high incident flux (>150\,$F_\oplus$) tend to have smaller radii.

The major reason for our limited sample size is due to the number of stars in the \textit{Kepler} field for which we had Str\"omgren photometry for. However, we anticipate that with the upcoming \textit{Gaia} DR3 release we can obtain a larger sample of e.g., Str\"omgren photometry (or other suitable metallicity sensitivity indices) directly from the $BP$ and $RP$ spectra. With this larger sample, we believe that the methods described in this paper will be limited solely by \textit{Kepler} uncertainties, thus allowing for more robust statistics and deeper insight into the demographics of \textit{Kepler}'s exoplanet population.

\section*{Acknowledgements}

L.C. is the recipient of the ARC Future Fellowship FT160100402.
M.I. acknowledges support from the ARC Discovery Scheme (DP170102233). This research has also made use of the NASA Exoplanet Archive, which is operated by the California Institute of Technology, under contract with the National Aeronautics and Space Administration under the Exoplanet Exploration Program.
This work has made use of data from the European Space Agency (ESA) mission \textit{Gaia} (\url{https://www.cosmos.esa.int/gaia}), processed by the \textit{Gaia} Data Processing and Analysis Consortium (DPAC, \url{https://www.cosmos.esa.int/web/gaia/dpac/consortium}). Funding for the DPAC has been provided by national institutions, in particular the institutions participating in the \textit{Gaia} Multilateral Agreement.

\section*{Data Availability}
 The data underlying this article are available in the article and in its online supplementary material.

%%%%%%%%%%%%%%%%%%%%%%%%%%%%%%%%%%%%%%%%%%%%%%%%%%

%%%%%%%%%%%%%%%%%%%% REFERENCES %%%%%%%%%%%%%%%%%%

% The best way to enter references is to use BibTeX:

\bibliographystyle{mnras}
\bibliography{main} % if your bibtex file is called example.bib

%%%%%%%%%%%%%%%%%%%%%%%%%%%%%%%%%%%%%%%%%%%%%%%%%%

%%%%%%%%%%%%%%%%% APPENDICES %%%%%%%%%%%%%%%%%%%%%

\appendix

%%%%%%%%%%%%%%%%%%%%%%%%%%%%%%%%%%%%%%%%%%%%%%%%%%

% Don't change these lines
\bsp	% typesetting comment
\label{lastpage}
\end{document}